# Photocurrent Characteristics of Individual GeSe$_2$ Nanobelt with Schottky Effects

Bablu Mukherjee, Eng Soon Tok and Chorng Haur Sow[*]

Department of Physics, 2 Science Drive 3, National University of Singapore (NUS), Singapore-117542

## Abstract

Single crystal GeSe$_2$ nanobelts were successfully grown using chemical vapor deposition techniques. The morphology and structure of the nanostructures were characterized using scanning electron microscopy (SEM), transmission electron microscopy (TEM), X-ray diffractometry (XRD) and Raman spectroscopy. Electronic transport properties, photoconductive characteristics and temperature-dependent electronic characteristics were examined on devices made of individual GeSe$_2$ nanobelt. The current increased by three orders of magnitude upon laser irradiation (wavelength 532 nm, intensity ~ 6.8 mW/cm$^2$) with responsivity of ~ 2764 A/W at fixed 4V bias. Localized photoconductivity study shows that the large photoresponse of the device primarily occurs at the metal-NB contact regions. In addition, the electrically Schottky nature of nanobelt/Au contact and p-type conductivity nature of GeSe$_2$ nanobelt are extracted from the current-voltage characteristics and spatially resolved photocurrent measurements. The high sensitivity and quick photoresponse in the visible wavelength range indicate potential applications of individual GeSe$_2$ nanobelt devices in realizing optoelectronic switches.

[*] Corresponding author. Email address: physowch@nus.edu.sg

# I. Introduction

The development of stable, fast and broad band individual nanobelt (NB) photodetectors has attracted interest towards optoelectronic device.[1,2] Previous work has demonstrated the key issue of nanostructures-metal electrode interface for nanoelectronic devices and described important factors that are responsible for the photoconductivity of individual NB devices. These factors include design of electrodes, contacts formation between nanowire-metal and the band bending effects caused by the surface states on the nanowire surface.[3-5] Single-NB devices enable careful study of fundamental processes such as charge transfer, surface recombination, and minority carrier diffusion.[6,7,8] The mechanisms of carrier generation with charge transport and collection in nanowire/nanotube photodetectors have been addressed.[9,10] Recently Varghese *et al.* have shown the ohmic nature of the nanowire/electrode contact and the p-type conductivity of individual nanowire from the current-voltage characteristics. In addition, spatially resolved photocurrent measurements was demonstrated.[11] Therefore it is important to quantify the important role of the contacts (nanostructures-electrodes) of the nanodevices and optoelectronic devices using non-destructive methods for better insights into their performance. It is also important to identify the possible reasons for response of the nanomaterial under external laser irradiation and find out the possible conduction mechanism of the photogenerated charge carriers in the nanodevice.

A large number of studies including steady-state photoconductivity measurements have been made of the amorphous germanium selenide system. Defect states in Ge based chalcogenides including $GeSe_2$ glasses have been observed by photoluminescence and ESR.[12] Two distinct localized states are available near to the mid band gap of $GeSe_2$. These states are interpreted as defect states with negative correlation energy, originated from Ge and Se atoms.[13] Recently, single-crystalline GeSe nanostructures including nanocomb and nanosheets have been reported for high-performance electronic devices based on layered p-type semiconductor. Germanium diselenide ($GeSe_2$) is also emerging as an important wide bandgap IV-VI semiconductors (Eg ~ 2.7 eV) with layered structure used in telecommunications applications and waveguides application.[14,15] Single-crystalline $GeSe_2$ nanowalls have been reported as a promising candidates for application in field emitter.[16]

In this work, we report the controlled synthesis of single-crystalline $GeSe_2$ nanobelts (NBs) by a simple chemical vapor deposition (CVD) method and demonstrate the electrical and photoconducting properties of individual $GeSe_2$ NB. Such high crystalline $GeSe_2$ NBs were synthesized at low temperature CVD process compared to recently reported $GeSe_2$ nanostructures.[16,17] We have discussed the possible growth mechanism of such catalytic growth of $GeSe_2$ NB. Using Raman spectroscopy at high laser intensity, local structural change from as-grown β-crystalline to α-crystalline phase have been observed in single $GeSe_2$ NB. Two terminal current-voltage (*I-V*) measurements and temperature dependent I-V measurements of single $GeSe_2$ NB device were performed. Temperature dependent I-V measurements indicate thermal effect as a possible transport mechanism. We performed photocurrent measurement of individual NB by irradiating the laser beam (wavelength 532 nm) uniformly over the NB device (global irradiation) and by applying spatially (~ 3 μm) resolved focused laser over the different parts of the individual NB device (localized irradiation). The current increased by three orders of magnitude upon laser irradiation. Localized photoconductivity study shows that the large photoresponse of the device primarily occurs at the metal-NB contact regions. High photoresponse of the individual $GeSe_2$ NB device is attributed to the thermal effect, surface states and electrical contacts of the device.

## II. Experimental Section

In our experiments, the synthetic route is described as follows. The mixture of pure Ge powder (Sigma Aldrich, purity 99.99%), Se powder (Sigma Aldrich, purity 99.99%) and carbon nanopowder (particle size < 50 nm, Sigma Aldrich, purity 99.99%) in molar ratio of 1:2:3 was used as source materials. The Au coated (15 nm thick) Si substrate and the small alumina boat containing a small amount (~ 0.4 gm) of as milled Ge:Se:C source powders were loaded into an one-end open quartz-glass tube. The tube was inserted in a horizontal quartz tube placed in a conventional tube furnace[18] such that the substrate was set at low-temperature region with respect to the mixed Ge:Se:C source powders and the distance between them was about 28 cm. Then the quartz tube was evacuated for 2 hrs by a vacuum pump and subsequently was filled with argon. The argon gas was allowed to flow for 1 hr. After that, the furnace was heated under an argon flow of 100 sccm (standard cubic centimeters per minute). When the temperature reached 680 $^o$C (heating rate: 30 $^o$C min$^{-1}$), the pressure of Ar carrier gas was maintained at ~ 2 mbar during synthesis for 30 mins. After the reaction was terminated, the substrate was coated with a layer of yellow product.

The morphology, structure and chemical composition of the as-synthesized nanostructures were characterized using field emission scanning electron microscopy (FESEM, JEOL JSM-6700F), transmission electron microscopy (TEM, JEOL, JEM-2010F, 200 kV), energy-dispersive X-ray spectroscopy (EDX) equipped in the TEM, X-ray diffraction (X'PERT MPD, Cu Kα (1.542 Å) and Raman spectroscopy (Renishaw system 2000, excitation 514.5 nm Ar$^+$ laser ). The Veeco D3000 NS49 AFM system is used to measure thickness of the NB.

$GeSe_2$ NB photodetectors were fabricated using standard photolithography technique. $GeSe_2$ NBs were first dispersed in ethanol and then dropped onto a $SiO_2$ (300 nm)/n-Si substrate. Standard photolithography and Au (thickness: 200 nm) film deposition and lift off were performed to pattern electrodes on the individual $GeSe_2$ NBs. The current-voltage (I-V) and all electrical characteristics of the $GeSe_2$ NB device was measured using Keithley 6430 source-measure unit with the device housed in a vacuum chamber. The photoresponse of these photodetectors was measured using continuous wave laser beam from the diode laser (532 nm, SUWTECH, LDC-2500). The laser beams could directly irradiate the NB device through a transparent glass window of the vacuum chamber and the device was electrically connected through a pair of vacuum compatible leads to the source meter. For localized photocurrent measurements, continuous wave laser beam from the diode laser (532 nm) was focused into a tiny spot (spot size ~ 3μm) using objective lens of an optical microscope. The test devices were placed onto the optical microscope sample stage and the position dependant localized photocurrent was measured by irradiating the focused laser on different position of the nanowire device. The devices were positioned on a heating stage for systematic variation in temperature and simultaneously the *I-V* characteristics were recorded with variation in temperature. The temperature of the device was controlled by a Linkam [TMS 94] temperature controller. All electrical and optoelectrical measurements of single NB based two probe devices were measured under vacuum of 0.001 mbar.

## III. Results and Discussions

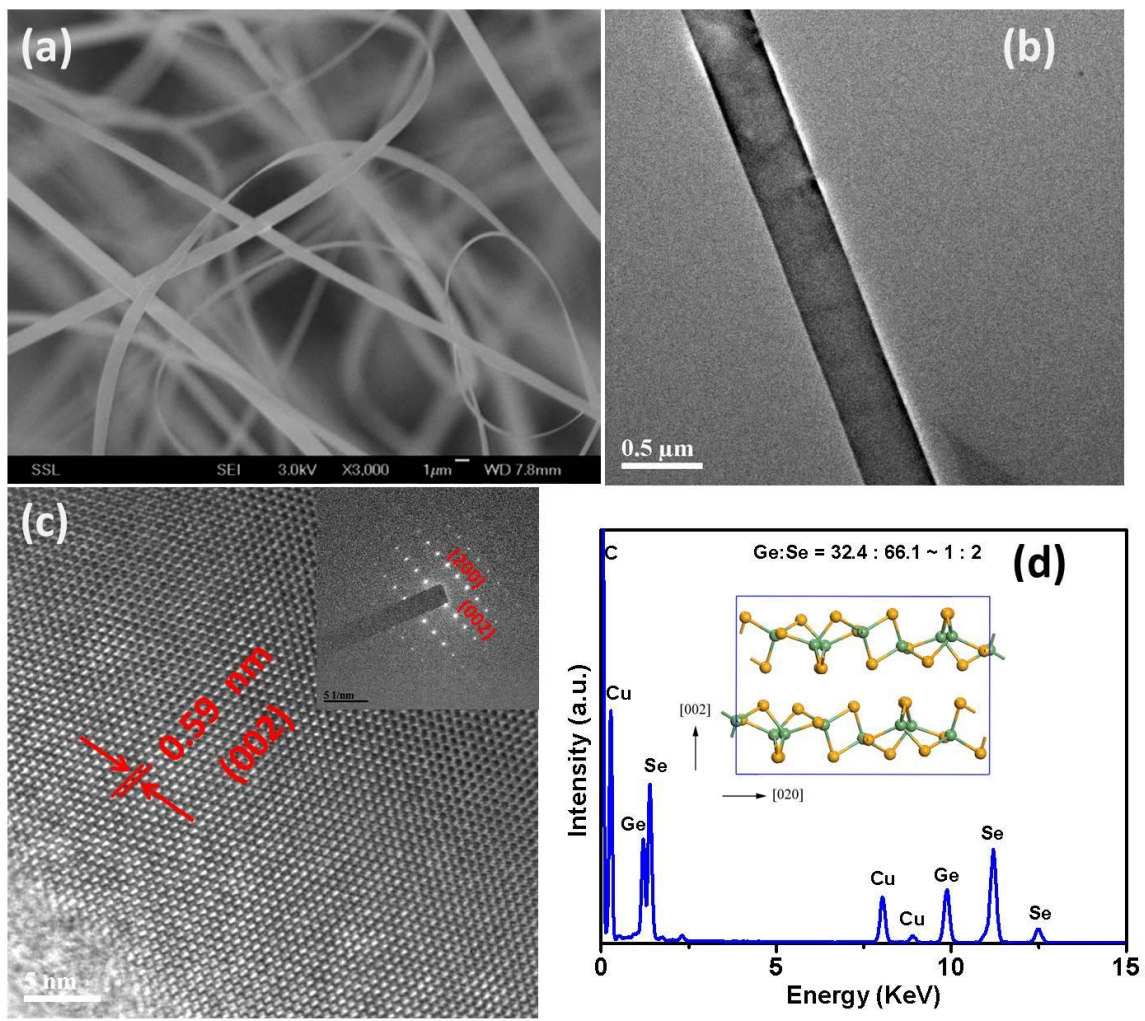

**Figure 1.** **(a)** FESEM image, **(b)** low-magnification TEM image, **(c)** HRTEM image, and **(d)** EDX spectrum of the as-synthesized GeSe$_2$ NBs. Inset in **(c)** shows the selected-area electron

diffraction (SAED) pattern. Inset in (**d**) shows the unit cell of GeSe$_2$ NB. Green balls and pale yellow balls represent Ge and Se, respectively.

The NB morphology is evident in **Figure 1a** in which a twisted belt with a thickness of several nanometer is portrayed. The FESEM image (**Figure 1a**) shows as-synthesized GeSe$_2$ NBs with an average width of ~ 2 μm and a length of ~ 50 μm. The thickness of the NBs is about 60-80 nm. The cross-section of the NB is rectangular, which can be seen from AFM image of a GeSe$_2$ NB.[19] The high-resolution TEM image confirmed that the GeSe$_2$ NB has a single crystalline nature. A low magnification TEM image of a single NB shows that the NB has smooth surfaces. The width and thickness of a given NB are uniform along the whole length. The GeSe$_2$ NBs grow along <002> direction, which gives rise to a relatively intense diffraction peak of the (002) plane. The high-resolution TEM image of a single NB (**Figure 1c**) exhibits clear fringes perpendicular to the NB axis. The SAED pattern of the single GeSe$_2$ NB in the [020] zone axis is inserted in **Figure 1c**, which further indicates its single crystalline monoclinic crystal structure with growth direction along [002] direction. The fringe spacing measures 0.59 nm, which concurs well with the interplaner spacing of (002) and alludes to the NB growth direction along [002]. This is consistent with the common growth direction of GeSe$_2$ NB along [002] direction, as seen in the growth direction of the stepped surfaced GeSe$_2$ NB.[20] On the basis of the XRD and TEM results obtained above, a structural model of the synthesized GeSe$_2$ NBs is inserted in **Figure 1d**. The inset shows the ball-stick model of the unit cell of GeSe$_2$ with layers along [002] direction, where green balls and pale yellow balls represent Ge and Se atoms, respectively.

The molecular ratio of Ge/Se calculated from the EDX data (**Figure 1d**) is approximately 1:2, which is in agreement with the stoichiometric ratio of GeSe$_2$ and confirms that the as-grown NBs are indeed composed of monoclinic structured GeSe$_2$ crystal.

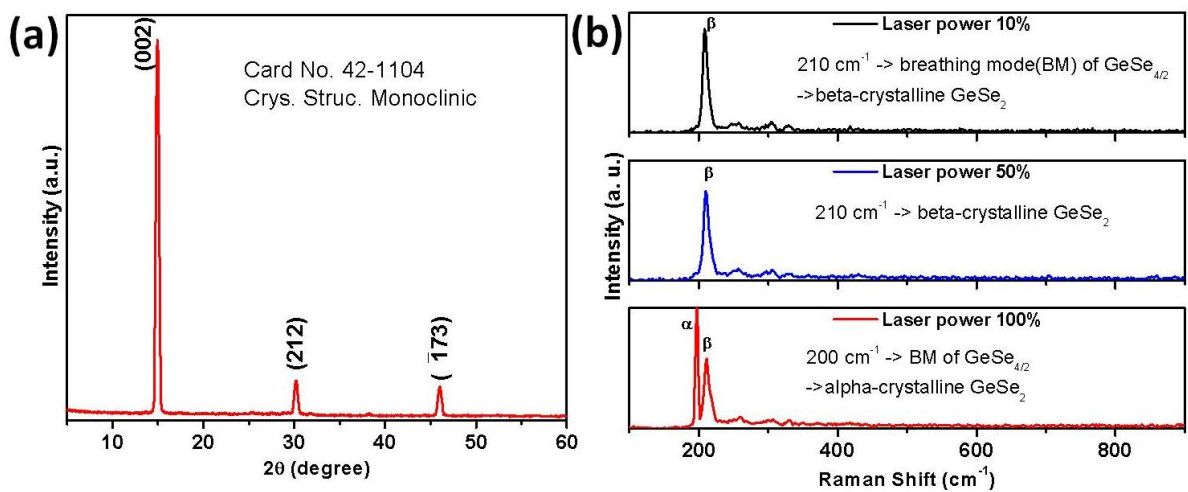

**Figure 2.** (**a**) Typical XRD pattern of GeSe$_2$ nanostructures. (**b**) Raman spectra taken during the laser induced crystallization process of crystalline GeSe$_2$ NBs. At full laser power, the NB

crystallizes into the α-phase (peak position at ~ 200 cm$^{-1}$) with accompanying β-phase (peak around ~ 210 cm$^{-1}$).

Shown in **Figure 2a** is the XRD pattern of the as-prepared NBs sample. It can be seen that the XRD pattern is in conformity with monoclinic GeSe$_2$ (a = 7.016 Å, b = 16.79 Å, c = 11.831 Å). In the XRD pattern of the NBs sample, the relative peak intensity of the diffraction plane (002) is much higher when compared with that of the standard powder diffraction pattern of bulk GeSe$_2$. **Figure 2b** shows the Raman spectra of as synthesized GeSe$_2$ NB with varying laser power excitation throughout the measurements. During the photo-induced crystallization process at around the laser excitation power of 100 %, we have observed two vibrational modes in Raman spectra. The Raman band located around 210 cm$^{-1}$ indicates β-form crystalline and a Raman band (at around 200 cm$^{-1}$) indicates α-form crystalline of GeSe$_2$ nanostructures.[21,22] Raman bands at 200 cm$^{-1}$ and 210 cm$^{-1}$ are assigned to the breathing mode (BM) of GeSe$_{4/2}$ tetrahedra with corner and edge sharing connection, respectively. The integrated intensity ratio of the α band and β band is ~ 1.5 : 1, which indicates the significant α-phase crystallization that occurred along with β-phase crystallization. One possible reason for the formation of α-phase crystallization along with β-phase crystallization is the local heating due to the irradiation of high laser power. However as shown in the other two Raman spectra in **Figure 2b** we have not observed any α-phase crystallization peak at low laser excitation power which confirms that pure β-crystalline GeSe$_2$ NBs were obtained during synthesis.[21,22] As our photocurrent measurements were performed at very low laser power irradiation, the photoresponce and optoelectronic properties were reflected of the nature of β-phase GeSe$_2$ NBs.

## A. Synthesis of GeSe$_2$ NBs

In the past few years, a number of reports addressed the growth mechanism of NBs of different material system.[23-25] Better understanding of growth process to obtain NBs of desired size and morphology is important in determining their properties. The most common synthesis method for NB system in vapor-phase synthesis is either VS growth[26] or catalyst-assist VLS growth.[27] The SEM and TEM images (**Figure 3 a,b**) of the grown tips of the NBs show that the diameter of the Au alloy catalyst is larger than the thickness of the produced NBs, which suggests that the thickness of the NBs can be controlled by the size of the Au catalysts.

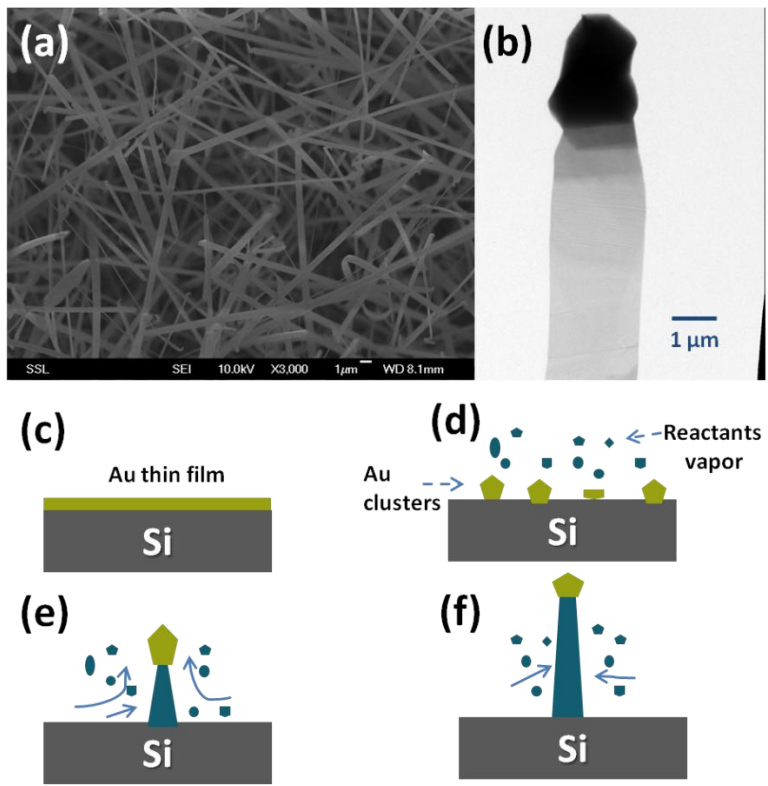

**Figure 3.** (a) Low magnification SEM image of the smooth surfaced GeSe$_2$ NBs. (b) Low magnification TEM image of the GeSe$_2$ NB with Au alloy clusters at the tip of the NB. Schematic illustrations of Au catalysts assisted VLS growth process of smooth surfaced GeSe$_2$ NB at high temperature: **(c-f)** show the possible stages of the NB synthesis.

It has been observed from the SEM image[28] of the growth substrate Si coated with thin seed layer of Au nanoparticles that each NB grows with the assistance of Au nanoclusters. Au alloy formation at the tip of the individual NB has been shown in the TEM image[28] and corresponding EDS elemental map confirms the presence of Au element[28] at the tip of single NB. By heating the source materials (mixture of Ge, Se powder with carbon) at a relatively high temperature of 680 °C, Ge and Se vapors are generated and transported to the downstream region with the help of inert Ar carrier gas. The liquid Au droplets, which are formed from the Au nanoparticles coated film on Si substrate at the elevated temperature,[29] act as nucleation sites and absorb the vaporized reactants (i.e. Ge and Se vapors). Continuous absorption, dissolution of Ge and Se in the Au-Ge-Se eutectic[20] alloy droplets will lead to saturation of GeSe$_2$, which results in the nucleation and growth of GeSe$_2$ NBs through VLS process. For longer time growth process, more reactance vapor of the source materials stay in the growth region for longer time, which assists longer growth of the NBs. This growth process is schematically shown in **Figure 3(c-f)**. This Au-catalyzed VLS growth process for NB synthesis is similar to those reported for the synthesis of In$_2$O$_3$ NBs,[30] Ga$_2$O$_3$ NBs[31] and CdS NBs.[32]

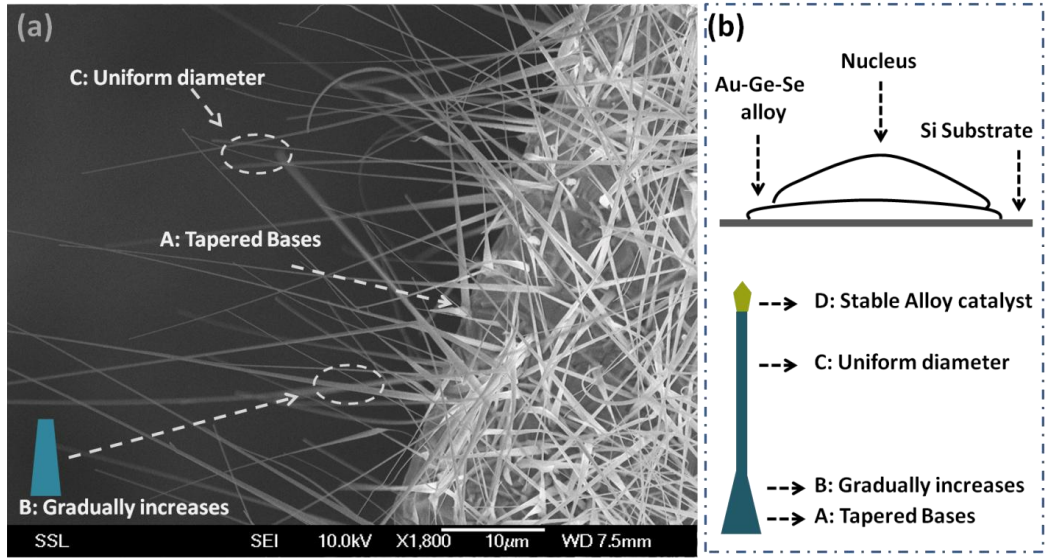

**Figure 4** **(a)** Low magnification SEM view of the GeSe$_2$ nanostructures grown from the edge of the substrates. The different parts of the grown nanostructure (i.e. tapered bases, gradually increments stem, and uniform diameter body) are marked in the SEM image. **(b)** (Top) Schematic of the nucleus formed on the molten Au-Ge-Se alloy droplet and growth of GeSe$_2$ NB at initial stage with tapering. (Bottom) Schematic representation of the grown NB specified with different regions.

Form the SEM image (**Figure 4a**); it has been observed that the morphology of the NBs is controlled by the growth of both axial and side growth. The side growth using VS growth process has been discussed in many nanostructures material system. Similarly this could be applicable in smooth surfaces GeSe$_2$ NBs that VLS growth process governs the growth along the belt length direction and VS growth process governs the side growth and belt thickness. The SEM image (**Figure 4a**) of the smooth surfaced GeSe$_2$ NBs on the edge of the growth substrate (Au coated Si substrate) shows that the bases of the NBs show tapering. The tapered segment of the NB is just restricted to the base, which is probably due to the limitation of adatom diffusion through the NB sidewall. It is reported that the upward adatom mobility via the NB sidewalls decreases at high temperature.[33] The tapering also indicates that the adatom surface diffusion from the substrate up the NB sidewalls forms a direct path for the growth species reaching the alloy droplet other than the possible direct impingement of the atoms on the droplet.[34] The tapering of the NB may explain the basis of stability and the shape of alloy catalyst, which was observed on the tip of the NB. **Figure 4b** (top) illustrates that the Au-Ge-Se alloy droplet form large contact area with Si substrate at the first stage and the diameter of the nucleus is controlled by the size of catalyst contact area at the liquid-solid interface. The contact angle measurements and surface tension at the interface of various regions at the initial stage of growth process with the modified Young's relation[35] may suggest that the rate of decreases in diameter of the nucleus is much larger close to base and then decreases as length increases. Thus the reduction of the size of the catalyst at the top of NB is observed and formation of the NB with stable alloy catalyst is observed too. **Figure 4b** (Bottom) shows the schematic representation of the grown NB, where the different regions of the NB are marked.

## B. Photocurrent measurements using broad beam irradiation

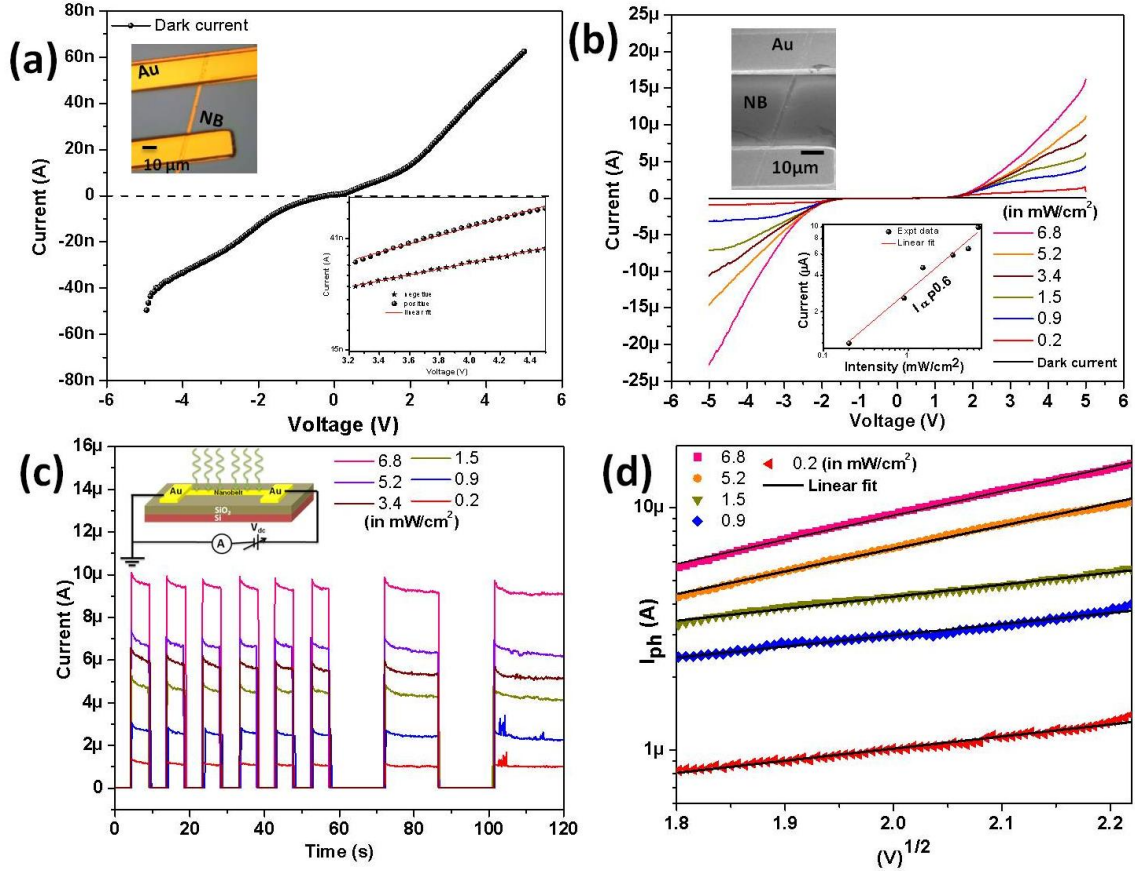

**Figure 5.** (a) Photoresponse of individual GeSe$_2$ NB devices under global laser illumination. Dark I-V curve recorded in sweeping bias of -5 V to +5 V. The insets show the optical image of the device (top) and the ln(I)-V curves with linear fit at intermediate voltage range (bottom). (b) I-V curves were recorded under dark condition and under global irradiation of 532 nm light with varying laser intensities from $0.2 \pm 0.1$ mW/cm$^2$ to $6.8 \pm 0.1$ mW/cm$^2$. Inset shows photocurrent dependency on laser intensity. Both photocurrent and light intensity are in the log scale. (c) Photocurrent-time (I-t) response (fixed dc bias: 4V) at the fixed wavelength excitation with varying intensities. The inset shows the schematic diagram of the device under global illumination. (d) Photocurrent ($I_{ph}$)-(voltage)$^{1/2}$ graphs with linear fit at intermediate voltage range with varying laser intensities.

Nanostructures based photodetectors have been extensively studied over the past decades.[36-40] Two-probe I-V and electrical measurements were carried out on individual GeSe$_2$ NB devices.

Photogenerated carriers could significantly improve the conductivity when the semiconducting nanomaterial is irradiated by photons with the energy higher than the band gap of the material.[41] For our experiments, we have selected laser beams with photon energy below band gap (laser wavelength: 532 nm) of the NB. **Figure 5a** shows typical *I-V* curves of a single GeSe$_2$ NB device, measured in the dark condition. Inset (top in **Figure 5a**) shows the optical image of the single GeSe$_2$ NB device. From the inset (bottom) of **Figure 5a**, we can see that ln(I) is linear with V in the intermediate bias voltage for both positive and negative currents, which indicates the typical I–V characteristic of back-to-back Schottky barriers (SB) structure.[42] The current across the NB significantly increased by three orders of magnitude, from ~ 5 nA (dark condition) to ~ 9 μA (Intensity ~ 6.8 mW/cm$^2$, 532 nm-light illumination) at applied bias of 4V.

The photoresponse of GeSe$_2$ NBs is dependent on laser intensity. **Figure 5b** shows the I-V curves of NB photodetector under dark conditions and after globally irradiated by 532 nm light with varying intensity. The photocurrent increases with the light intensity, consistent with the fact that the charge carriers' photogeneration efficiency is proportional to the absorbed photon flux. The corresponding dependency is plotted in log-log plot as inset in **Figure 5b**. This can be fitted to a power law, I$_{ph}$ ~ P$^\gamma$, where γ determines the response of the photocurrent to the light intensity and photocurrent I$_{ph}$ = (I$_l$-I$_d$) and I$_l$ is the current of the device under illumination with the light source and I$_d$ is the dark current. The fitting gives a linear behavior with γ = 0.6. The non-unity (0.5<γ<1) exponent suggests a complex process of electron-hole generation, recombination, and trapping within a semiconductor.[43,44] As the 532nm light was illuminated over the full devices, the different parts of the device will contributes to the IV characteristics of the device, which is further supported by the localized photocurrent experiments.[11] **Figure 5c** shows the reversible switching of the NB photodetector between low and high conductance states when 532 nm light was turn on and off with varying laser intensity. The photogenerated current increases with increment of laser power at fixed dc bias. It is noted that the response time of such photodetector is quick (rise time: ~ 0.1s and decay time: ~ 0.1s), which is independent of laser intensity. The devices were prompt in generating photocurrent with a reproducible response to ON-OFF cycles. From the time response curves (I-t curves), we have noticed that sharp fall in photocurrent to dark current upon blocking of the laser irradiation to the device, which suggests that the photogenerated carriers might not follow the mechanism of discharging the NB surface adsorption molecules to contribute into the photocurrent.[45] Spectra responsivity (R$_\lambda$) and external quantum efficiency (EQE) can be calculated from the equations: $R_\lambda = \frac{I_{ph}}{P_\lambda S}$, and $EQE = \frac{h\,C\,R_\lambda}{e\,\lambda}$,[46,47] where I$_{ph}$ is the photocurrent, P$_\lambda$ is the light intensity, S is the effective illumination area of the device, λ is the wavelength of the exciting laser wavelength and c is the velocity of light, and e is the electronic charge. The calculated R$_\lambda$ and EQE for the single NB device are ~ 2764 A/W and ~ 6.4 × 10$^5$ %, respectively, for the incident wavelength of 532 nm at 4 V, which are higher than those reported chalcogenides semiconducting nanophotodetectors.[48-50]

The nonlinear curves under illumination (**Figure 5b**) indicate the photocurrent response might be dominated by SBs formation at the contacts. In this NB device as a Metal-Semiconductor-Metal (MSM) structure, both the forward and reverse biased current increases under global laser light illumination. This implies photoexcited electron-hole (e-h) pairs significantly increase the concentration of majority carriers of the semiconductor in MSM structure, where the SBs heights also modulate with the light illumination.[51,52] Here we adopt a similar strategy in analysis, where the approximate photocurrent equation of ZnO nanowire based Schottky photodiode can be described as: $\ln(I_{ph}) \alpha V^{1/2}$.[53] **Figure 5(d)** shows the plot of $I_{ph}$ with $V^{1/2}$ in semi-log plot with different intensities of 532 nm light irradiation and all curves are fitted with linear equation. These results indicate that our photocurrent responses of the devices are consistent with Schottky contacts dominated photoresponse of MSM structure.

## C. Photocurrent measurements by localized laser irradiation

Spatially resolved and localized photocurrent measurements on $GeSe_2$ NB device were performed. The photocurrent-time (I-t) response was recorded from three different positions of the device while the device was not biased. **Figure 6a** shows optical images of the NB device captured during the photocurrent measurements with focused laser beam (532 nm) irradiating at three different positions. At zero bias photocurrents of opposite polarity were generated upon irradiation of focused laser beam near NB-electrode contact regions as displayed in **Figure 6a**. We obtained a smaller photocurrent (**Figure 6b**) when the zero biased device was irradiated with focused laser beam near to the midpoint of the NB. To study the effect of applied bias, I-V characteristics of the device were recorded by irradiating the focused laser beam (532 nm, fixed laser power ~ 14.2 ± 0.1 mW) at three different position of the device separately as shown in **Figure 6c**.

Under localized illumination via a focused laser beam (excitation spot size is nearly equal to 3 µm), the measured response of the device will reflect only the part of the device that was illuminated by the focused laser beam. Unlike the symmetrical $I_{ph}$-$V$ curve observed for global irradiation, the local $I_{ph}$-$V$ curves are asymmetrical and strongly positional dependent. Specifically, the photocurrent was largest when the illuminated region corresponds to the metal electrode-NB contact region. This implies that the photogenerated carriers' collection under global illumination may not be occurring uniformly across the entire device and the efficient separation of photogenerated charge carriers occurs at the contacts.

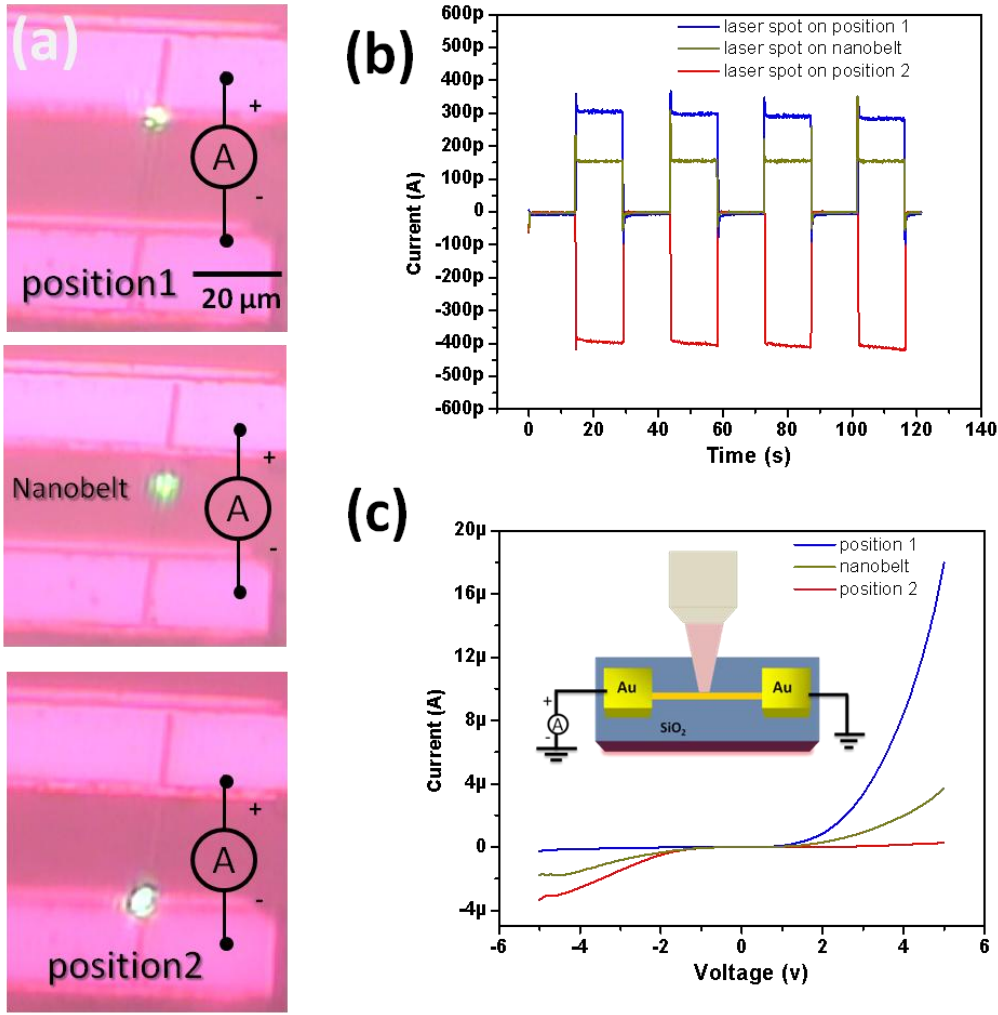

**Figure 6.** (a) Optical images of the GeSe$_2$ NB device with focused laser beam (bright spot) irradiated at its different position. (b) Photocurrent-time response upon periodic irradiation of focused laser beam (fixed power: 14.2 ± 0.1 mW) on three different positions of the NB device at zero bias voltage. (c) I-V characteristics of the NB device with focused laser beam on different positions. Inset shows the schematic of the NB device when localized laser is illuminating the middle of NB.

**Figure 6c** shows $I_{ph}$-$V$ characteristics under fixed localized laser power of ~ 14.2 ± 0.1 mW, which indicates the highest positive bias current was obtained when the forward (positive biased) metal electrode-NB junction was illuminated whereas the highest negative bias current was obtained when the reverse (negative biased) metal electrode-NB junction was illuminated. The asymmetrical intermediate $I_{ph}$-$V$ characteristic was obtained under the localized laser beam irradiation of the middle part of the device. At positive sweep bias, the drop in resistance of the NB device under localized laser irradiation on Position 1(biased electrode) is significantly higher than the localized laser irradiation on midpoint of the NB and Position 2 (grounded electrode).

Similarly at negative sweep bias, the drop in resistance of the NB device under localized laser irradiation on Position 2 (grounded electrode) is significantly higher than the localized laser irradiation on midpoint of the NB and Position 1 (biased electrode), which supports the scenario of majority carriers diffusion as described above.

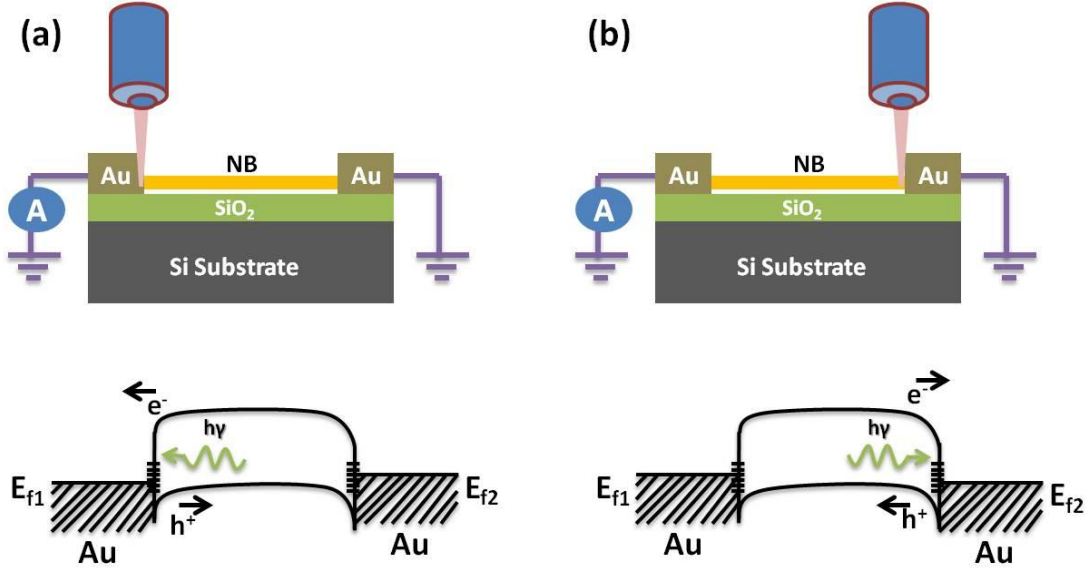

**Figure 7. (a,b)** Schematic diagram of focused laser at two ends of NB-Au junctions with their corresponding band diagram at zero bias condition ($E_{f1}$ and $E_{f2}$ are modified due to thermalization upon laser irradiation).

Considering the NB device as a MSM structure consisting of two Schottky barriers in back-to-back arrangement, it is observed that the photogenerated carriers, which are created by localized laser irradiation, can be efficiently separated from each other by the strong local electric field near the M-S interface.[11,54] Thus the increase in photogenerated charge carrier density upon irradiation can modify the barrier width, resulting in a narrow Schottky barrier at the contacts. This may facilitate an increase in the tunneling of photogenerated carriers from NB to metal electrodes through the modified Schottky barrier, which can be understood from energy band diagram **Figure 7(a,b)**. In addition, the localized thermal heating within the focused laser irradiated region resulted in thermoelectric effects at the M-S interface.[55] Under localized laser irradiation at the midpoint of the NB, the photogenerated carriers can diffuse randomly and recombine efficiently due to absence of strong electric field in the middle of NB, which generates low amount net photocurrent. Hence the polarity of the photocurrent measurements at zero bias on different positions of the device support such a scenario of hole diffusion. Thus the direction of the movement of holes is opposite when laser is shining at the different contacts at zero bias condition. Different contacts and Au-NB coupling lead to different magnitudes of photocurrent at the two contacts.

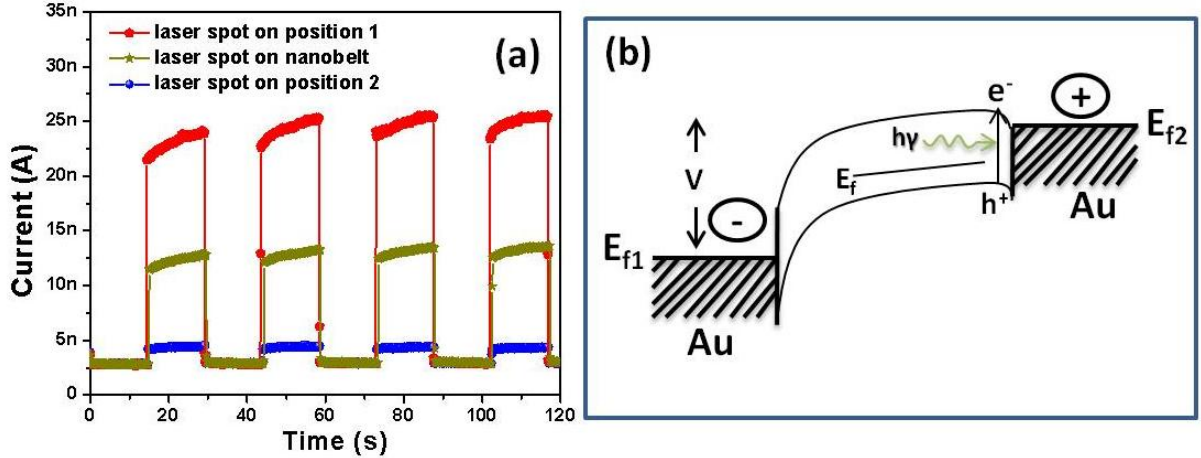

**Figure 8.** **(a)** Rising and decaying photocurrent characteristics upon periodic irradiation of focused laser beam on different portions of the NB device at a fixed bias of +1 V. **(b)** Schematic energy band diagram of the MSM structure at the applied positive bias condition.

**Figure 8(a)** shows that the rising and decaying photocurrent characteristics correspond to periodic irradiation of focused laser beam (532 nm; 14.2 ± 0.1 mW; at a fixed bias of +1 V) at different portions of the NB device. Different from the zero bias case, the laser irradiation near the midpoint of the NB produced significant photocurrent at a fixed bias of 1 V. When there is no external driving force at zero bias, the photo-generated carriers near the midpoint of the nanowire diffuse randomly rendering a net zero current. Whereas at a finite bias, the photogenerated carriers are drifted by the external electric field, and a corresponding photocurrent is recorded by the ammeter.

For forward bias voltage (positive bias) and shining of focused laser beam at position 1 (the contact with high electrostatic potential point), there is a much larger current. In this scenario 1, the Fermi level of bulk metal of right electrode (energy level: $E_{f2}$) is raised in energy, which lowers the potential drop across the depletion layer to a value by the amount of the reverse bias and increase hole current from the semiconductor to the metal. The energy band diagram of the biased MSM structure[56] is shown in **Figure 8b**. The holes, the dominant carriers in p-type GeSe$_2$, produced are drifted to the right side (since the holes tend to move to high electronic levels) whereas electrons move to the left end to contribute in net current and localized photogenerated electron-hole pair can be effectively separated out in the position 1, which gives rise to the observed large positive photocurrent. Under same external bias (+1 V to $E_{f2}$, position 1) configuration and shining of focused laser beam at position 2, there is lower photocurrent as compared to position 1. In this scenario 2, the barrier height for holes is increased on the semiconductor side by the amount of the reverse bias at position 2. The drift of holes from semiconductor to left metal electrode (energy level: $E_{f1}$) is therefore decreased at position 2. Thus localized photogenerated electron-hole pair cannot be effectively separated out in the position 2, which eventually produce lower observed photocurrent as compared to position 1. The lowest current obtained at position 2 may be due to the lower mobility of electrons. As the electrons need to drift from left side to right side along the whole belt, due to this low mobility, most of the electrons are scattered and the current is low. This also explains that why obtain a larger current at the middle position. In the middle position, the path is around a half compared to the case of position 2. Therefore the current for shining at middle is larger than shining at position 2. Under same external bias configuration and shining of focused laser beam at the NB body part, in this scenario 3, the photogenerated electron-hole pair can be effectively separated out, where the holes will drift to right electrode and electrons will drift to the left electrode as GeSe$_2$ is a p-type material, the mobility of electrons is low and most of electrons are scattered and cannot reach the left electrode, which produce intermediate photocurrent as observed in **Figure 8. (a)**.

## D. Temperature dependent I-V characteristics

To gain further insight into the charge conduction in the NB, we have performed the controlled experiments of temperature dependent dark dc conductivity. **Figure 9a** shows the temperature-dependent *I-V* curves of the single NB device in the standard two-probe measurement configuration. The symmetrical non-linear *I-V* characteristics were maintained with increased temperature.

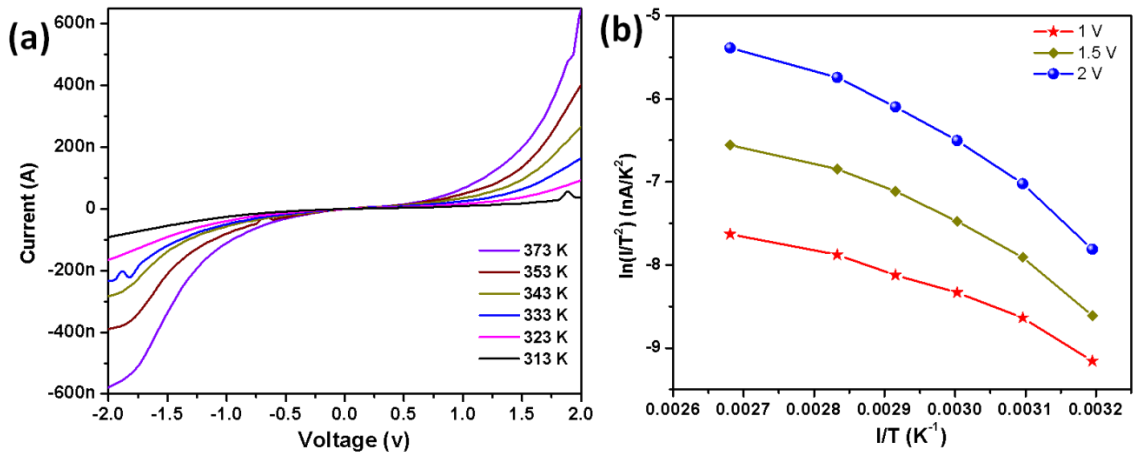

**Figure 9.** (a) I-V curves of individual GeSe$_2$ NB device at a temperature range from 313 K to 373 K. The scanned voltage range was from -2 V to +2 V. Inset at bottom right shows temperature dependence of conductivity of the single GeSe$_2$ NB device. (b) The $\ln(I/T^2)-(1/T)$ curves at various biases of 1, 1.5, and 2 V.

We have recorded the I-V curves at various temperatures by thermally anchoring the SiO$_2$/Si wafer with NB devices onto a heating stage. The dark IV curves (**Figure 9a**) at higher temperatures show non-linear and nearly symmetrical behavior, which preserve the dark IV characteristic at room temperature. The conductivity at higher applied bias range increases significantly with increased higher temperature. We have estimated $\ln(I/T^2)$-$(1/T)$ curves at different fixed bias as shown in **Figure 9b**. Such non-linear behaviors of $\ln(I/T^2)$-$(1/T)$ curves could be due to the thermal effect and the desorption process on the surface of NB.[57] As the temperature of the NB increases, the desorption of ion species such as oxygen ions ($O_2^-$, $O^-$, etc.) and $OH^-$ becomes easier. Thus the thermal effect and the surface traps states have considerable effect on the transport characteristics of the NB device.

The localized photocurrent experiments show the generation of higher photocurrent occurs at the interface of NB-electrode Schottky contacts. Temperature dependent IV characteristics and laser power dependency (power law) on the photocurrent under global irradiation indicates the thermal effect on the NB and surface traps states of the gaseous ions have considerable effect on carriers' transport. The time response curves (I-t curves) suggest that the photogenerated carriers might not follow the mechanism of discharging the NB surface adsorption molecules to contribute into the photocurrent. The photocurrent equation $\ln(I_{ph}) \alpha (V)^{1/2}$ plots shows Schottky contacts dominated photoresponse of MSM structure. The optoelectronic characteristics of the individual $GeSe_2$ NB device indicate that the thermal effect, surface traps, and SBs at contacts are the dominant mechanism for the photoresponse.

## IV. Conclusions

In conclusion, the NBs of $GeSe_2$ were synthesized in a horizontal tube-furnace. XRD, Raman spectroscopy, SEM and TEM were used to characterize and investigate the morphology and structure of the nanostructures. Two electrode pads were deposited on the ends of individual NB to form simple devices with the formation of SBs at the contacts. The electrical and optoelectrical characteristics of individual $GeSe_2$ NB devices were studied. A high photoresponsivity of 2764 A/W and quick response time of ~0.1s were observed from single $GeSe_2$ NB devices under 532 nm-light irradiation. The temperature dependence of current-voltage curve of single $GeSe_2$ NB devices was studied. Localized photocurrent study and optoelectronic characteristics under global laser irradiation study provide valuable insight into the carrier transport in the individual NB devices. Although there are many factors which rules the higher photoresponse of the device, high photoresponse of the individual $GeSe_2$ NB device indicate that the photoresponse is attributed to the thermal effect, surface traps, and SBs at the electrical contacts.

# Supporting Information

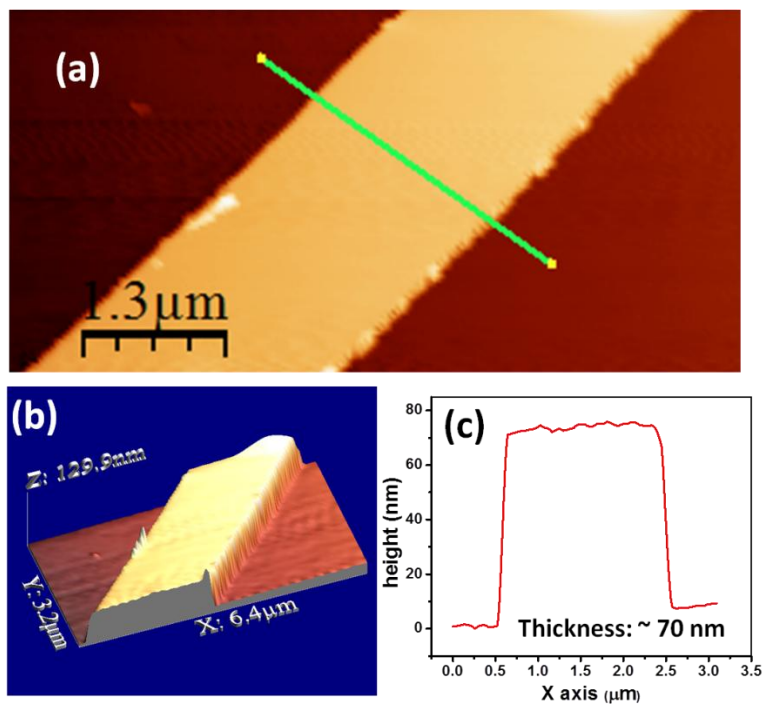

**Figure S1.** (a) AFM image of a GeSe$_2$ NB. (b) 3D AFM image of a single GeSe$_2$ NB. (c) Cross sectional profile, which is drawn along the marked line in (a), shows the thickness of the NB is 70 nm.

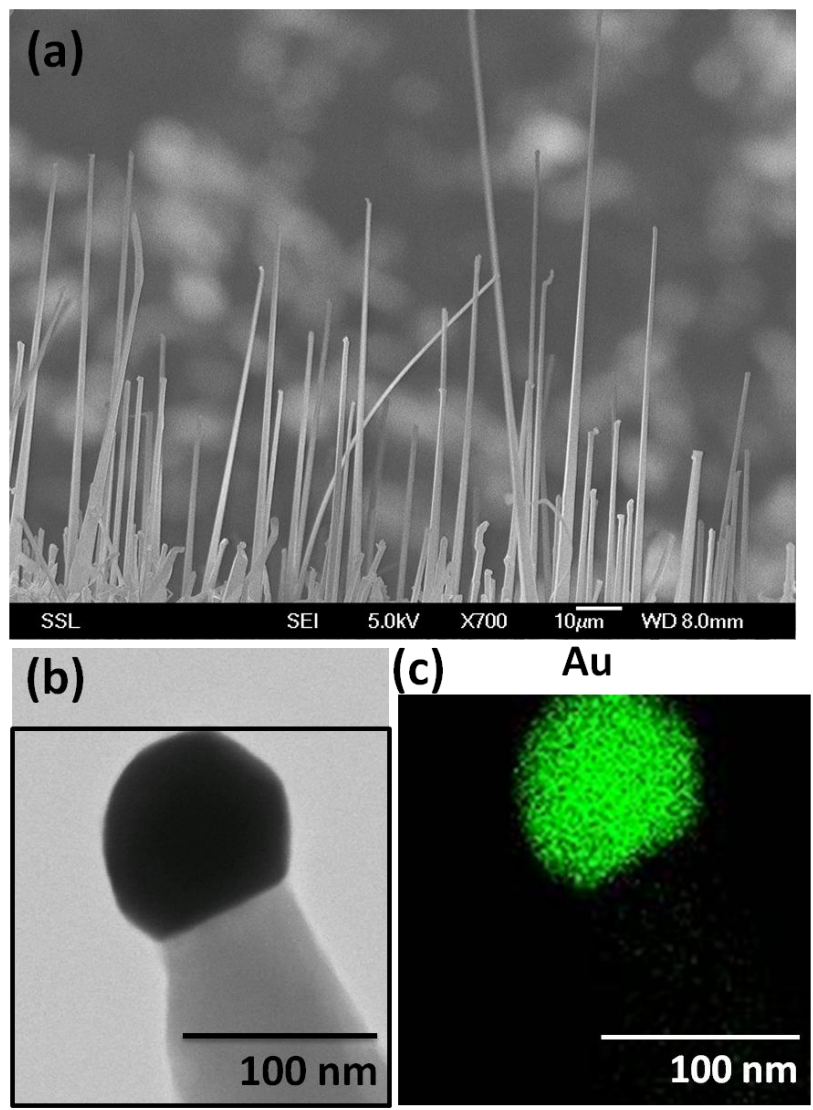

**Figure S2.** **(a)** SEM image of GeSe$_2$ NBs grown from the edge of Au film coated Si substrate. **(b)** TEM image of the Au cluster formation at the tip of the single NB. **(c)** false-color energy dispersive X-ray spectroscopy (EDS) elemental map of Au in the rectangular region defined in **(b)**.